# A steady state approach for studying valley relaxation using optical vortex beam

Aswini Kumar Pattanayak, Pritam Das, Avijit Dhara, Devarshi Chakrabarty, Shreya Paul, Kamal Gurnani, Maruthi Manoj Brundavanam, Sajal Dhara[*]

*Department of Physics, Indian Institute of Technology, Kharagpur-721302, India*

*\* sajaldhara@phy.iitkgp.ac.in*

**Abstract:**

Spin-valley coupling in monolayer transition metal dichalcogenides gives rise to valley polarization and coherence effect, limited by intervalley scattering caused by exciton-phonon, exciton-impurity, and electron-hole exchange interaction (EHEI). We explore an approach to tune the EHEI by controlling excitons center of mass momentum (COM) utilizing the photon distribution of higher-order optical vortex beam. By virtue of this, we have observed excitons COM-dependent valley depolarization and decoherence which gives us the ability to probe the valley relaxation timescale in a steady-state measurement. Our steady-state technique to probe the valley dynamics can open up a new paradigm to explore the physics of excitons in two-dimensional systems.

Probing valley dynamics in 2D materials is indispensable and has been in the spotlight of current research addressing challenges towards the utilization of valley degree of freedom (DoF) in quantum information processing and valleytronics[1–5]. A finite value of the degree of valley polarization (DoP)[6–8] and degree of valley coherence (DoC)[9,10] is observed in steady-state measurement, limited by intervalley scattering caused by exciton-phonon, exciton-impurity[7,11], and more importantly by the long-range electron-hole exchange interactions[6,12] (EHEI). Studying the time scale associated with these valley exciton dynamics is important for the realization of devices employing the valley degree of freedom. These time scales associated with exciton dynamics have been probed before via time-resolved PL[10,13,14], absorption[15], Faraday, and Kerr rotation techniques[16]. However, probing such dynamics to identify intervalley EHEI scattering contributions in a steady-state measurement scheme to our knowledge has not been explored so far.

The mechanism of valley dynamics at low temperature is dominated by the long-range EHEI, which largely depends on excitons center of mass momentum (COM)[17]. Whereas at high-temperature regime it is dominated by exciton-phonon scattering process[18]. In conventional PL spectroscopy using a gaussian beam, it is difficult to tune excitons COM to study the effect of momentum change associated with the scattering process. Here, we have proposed an approach to probe the valley dynamics in TMDs by tuning the COM of the exciton, by using an optical vortex beam (OV) or twisted light beam[19,20]. OV beams have drawn great attention for the last three decades for carrying orbital angular momentum (OAM) $l\hbar$ per photon along with spin angular momentum[21,22] due to their helical phase profile ($e^{-il\phi}$). OV beam find myriad applications, like in optical trapping, super-resolution fluorescence microscopy, quantum communication and radio communication[19,23]. Recently we have demonstated tunable PL emission in $WS_2$ and $WSe_2$ monolayer by controlling intravalley scattering of excitons and longitudinal electric field component of light with excitation by twisted light[24].

In the resonant excitation process, the COM of bright exciton in TMDs monolayer is determined by the in-plane momentum of the exciting photon, which we normally neglect in the photoluminescence process. But intervalley scattering rates depend quadratically on these excitons COM,[12,17,25] received from the small spread of in-plane photon momentum within the light cone under the resonant excitation. The in-plane momentum of the light can be controlled by changing the $l$ of the OAM beam, which has been reported already [24,26]. Here we evince the interaction of OV beams with the excitons in the WSe$_2$ monolayer, by observing the valley depolarization and valley decoherence properties. The intervalley EHEI between excitons at K and K′ valleys can be probed via variation of DoP and DoC at a fixed temperature (T) with an external $l$ parameter affecting the scattering process. By keeping T fixed for a given sample, the phonon and impurity contributions[7,27] can be eliminated from the variation of the optical response with respect to the external parameter.

The schematic of the WSe$_2$ monolayer being excited by the OV beam through an objective lens of numerical aperture 0.7 has been presented in Figure 1a. The excitation of 660 nm laser is chosen which is in near resonant with the bright A excitons and the power has been kept below 300 nW before the objective to avoid sample heating. The OV beam of different $l$ is produced by an amplitude-only spatial light modulator displaying respective computer-generated holograms. For our work, we have considered the OV beam to be a Laguerre Gaussian beam (LG). The Fourier distribution of the LG beam at the focal plane for different $l$ numbers has been shown in our recent work[24]. It can be observed that the root mean square in-plane momentum of the LG beam is shifting towards higher momentum regime with the increase in $l$. In the excitation process, the in-plane momentum of the photon transfers into the momentum of the exciton due to the momentum conservation principle, so the root mean square momentum of photons is considered as excitons COM ($q_{r.m.s}$) as presented in Figure 1b. With increase in $q_{r.m.s}$ the intervalley scattering rate increases ($\gamma_{KK'} \propto q^2$) [12,17,28] with $l$, as shown by the dotted arrows in Figure 1b. Intervalley scattering, which plays an important role in determining DoP and DoC, is associated with e-h exchange interaction, phonon scattering, and impurity scattering. Among different intervalley scattering processes as described above, the long-range part of the EHEI[18] is dominant at low T-regime. This EHEI process is schematically shown in Figure 1c. Excitonic state at a particular valley ($K$ or $K'$) can be generated coherently by circularly polarized light excitations ($\sigma_+$ or $\sigma_-$) at near resonant condition with valley relaxation timescale more than the radiative recombination time[2,7,9]. By EHEI scattering mechanism, excitons change their valley degree of freedom[12] via simultaneous recombination of excitons in the $K$ valley and generation in the $K'$ valley, or vice versa. The effect of excitons motion on intervalley scattering is probed by measuring $l$ dependent DoP and DoC, which demands a highly sensitive polarization-resolved setup that avoids any unwanted change in the excitation laser and PL polarization state due to reflections in the optical path. For sensitive polarization-resolved measurements, special care has been taken to design a setup as shown in Figure 2 to avoid any kind of unwanted change in the relative phase between the $p$ and $s$ polarized components under reflections in the optical path. The complex reflection coefficients for $p$ and $s$ polarized components, $r_p$ and $r_s$, are in general unequal in phase and magnitude, resulting in the change of state of polarization under reflection. A polarized excitation beam with electric field components $E_p$ (along the horizontal plane of incidence) and $E_s$ (component perpendicular to the plane of incidence) undergoes reflection (at $45^0$ angle) by the beam splitter BS-1 with amplitudes $r_p E_p$, and $r_s E_s$ respectively followed by a 2nd reflection by an identical BS-2 where the plane of incidence is now vertical and hence interchanges the role of $p$ and $s$ polarization. Therefore, the final electric field components after two successive reflections are given by $r_s r_p E_p$, and $r_p r_s E_s$ respectively, hence keeping the state of polarization preserved. The exact configuration is adopted for the optical path of the PL signal using two identical mirrors. The polarization-resolved setup has been characterized with a controlled experiment with zero degrees of valley polarization at 295 K and coherence at 200 K as shown in Figure S3.

A $\sigma_+$ excitation beam is used to measure DoP, $\rho_c = \frac{I(\sigma_+)-I(\sigma_-)}{I(\sigma_+)+I(\sigma_-)}$, where, $I(\sigma_+)$ and $I(\sigma_-)$ are the intensities of $\sigma_+$ and $\sigma_-$ components of emitted PL intensity. Polarization resolved PL spectra are shown in Figure 3a, which are normalized with respect to $\sigma_+$ component of PL intensity for each $l$ at 4 K. The peak appearing around 1.747 eV has been identified as bright A exciton peak[9,29]. An eye guide is used to indicate the increase of $\sigma_-$ component in Figure 3a, demonstrating the depolarization effect of excitons with increasing $l$. The DoP is 24%, using a typical Gaussian beam ($l = 0$) at $T = 4$ K, which drops to 20% with increasing $q$ (corresponding $l = 0$ to $l = 6$) as shown in Figure 3b. Valley depolarization data at 30 K are shown in the Supporting Information Figure S5 where DoP decreases from 16% to 12% with increasing $q$. As with change in $l$ number of OV beam the spatial profile of the beam is changed, we have measured DoP in close vicinity points to address variation spatially in the sample. We have found the sample is showing almost homogeneous DoP of $(24.5 \pm 1)\%$ at four points as shown in Supporting Information S4. And we have also observed a similar kind of depolarization effect with $l$ number at different positions. This suggests the observed effects are solely coming due to the momentum change of exciton induced by the OV beam.

Similarly, a linearly polarized excitation beam is used to measure the DoC, $\rho_l = \frac{I_\parallel - I_\perp}{I_\parallel + I_\perp}$, where, $I_\parallel$ and $I_\perp$ are measured PL intensity in a pass and cross-orientation of the polarizer. PL spectrum for the pass and cross-orientation of the polarizer is shown in Figure 4a, which is normalized with respect to $I_\parallel$ component of PL intensity for each $l$ at 4 K. An eye guide is used to indicate the increase of cross-polarized PL components as shown in Figure 4a, demonstrating the decoherence effect with increasing $l$. The DoC is 56% using a typical Gaussian beam ($l = 0$) at $T = 4$ K, which drops to 50% with increasing $q$ (corresponding $l = 0$ to $l = 6$) as shown in

Figure 4b. Valley decoherence data at 30 K are shown in the (Supporting Information Figure S6). Where DoC decreased from 24% to 20% with increasing $q$.

To understand the physical process associated with depolarization and decoherence of valley exciton we have considered the Maialle-Silva-Sham mechanism[10,12,25]. This mechanism involves e-h exchange interaction, which is coupled with excitons momentum (q). This exchange interaction acts as a momentum-dependent in-plane effective pseudo magnetic field $\Omega(q)$ as shown in Figure 5a. The scattering of exciton momentum state is responsible for a fluctuating $\Omega(q)$ leading to the relaxation of pseudospin. Excitons momentum is increasing with $l$ as indicated by the concentric circle (dotted) in momentum space, increasing the magnitude of $\Omega(q)$ linearly with q as indicated by vector arrow (blue). This leads to an increase in excitons pseudospin precession frequency with $l$ (as shown in red, yellow, and blue cones for $l = 0 - 2$) around $\Omega(q)$ at higher momentum states. The relaxation of the longitudinal component of the valley pseudospin leads to the valley depolarization effect, associated with the relaxation time of $\gamma_{KK'}^{-1}$, whereas transverse component relaxation leads to the valley decoherence effect, associated with a relaxation time of $\gamma_{dep}^{-1}$. Both components of pseudospin relaxation rates quadratically depend on exciton momentum[12,28], which can be increased with increasing $l$ leading to depolarization and decoherence effect. To extract the relaxation time associated with valley dynamics, we have used rate equation models[6,24,30,31] to fit COM-dependent DoP and DoC as explained in Supporting Information S7. Intervalley spin-non-flip scattering[32,33] of electrons in conduction band can also be a possible pathway leading to valley depolarization. However, such spin-non-flip process involves electron-phonon interactions[34] which can be treated as a constant background for a fixed bath temperature and hence has been excluded in our analysis for simplicity. The DoP obtained from the rate equation model can be written as, $\rho_c = \frac{1}{1+\frac{2\gamma_{KK'}}{\gamma_b+\gamma_{bd}}}$, where $\gamma_b$, and $\gamma_{bd}$ are the radiative recombination, and intravalley scattering ($X^b \rightarrow X^d$) rate respectively. As both $\gamma_{KK'}$, and $\gamma_{bd}$ quadratically dependent on $q$, DoP can be fitted with, $\rho_c(q) = \frac{1}{1+\frac{2bq^2}{aq^2+\gamma_b}}$ (where, $a, b$ proportionality constants, see Supporting Information S7), as shown in Figure 3b. The valley decoherence has been fitted using the expression[31], $\rho_l = \frac{1}{\left(1+\frac{\gamma_v+\gamma_{dep}}{\gamma_x}\right)}$, where, $\gamma_x, \gamma_v$ are the exciton relaxation, and incoherent intervalley scattering rate respectively. As $\gamma_{dep} \propto q^2$, the DoC is fitted with the expression, $\rho_l = \frac{1}{1+\frac{\gamma_v+cq^2}{\gamma_x}}$, where, $c$ is a proportionality constant (See Supporting Information S7) as shown in Figure 4b. From fitting, we obtained valley polarization and valley coherence relaxation time scales for 4 K and 30 K as a function of $q$ are shown in Figure 5b and 5c respectively. The valley polarization timescale is found to be order of ~ 2 ps (at 4 K) and ~ 1 ps (at 30 K) which are in fine agreement with earlier reported time-resolved measurements[35,36]. We have obtained sub-picosecond valley polarization timescale for higher COM of exciton, whereas ~60 ps (at 4 K) and ~9 ps (at 30 K) valley coherence timescale has been obtained for monolayer WSe$_2$. From fitting parameters, the obtained intravalley scattering rate with $q$ has been shown in Supporting Information Figure S8.

In conclusion, we have utilized tunable dispersion of OV beam as a key to control exciton COM and demonstrate a steady-state measurement approach to probe the valley dynamics in monolayer WSe$_2$. With the increase in excitons COM the magnitude of the effective pseudo magnetic field is increased, which increases the precession frequency of pseudospin giving rise to valley depolarization and decoherence effect with higher-order $l$ of OV beam. At a fixed temperature, contributions of pseudospin relaxation due to exciton-phonon, and impurity scatterings are treated as a constant background without affecting our results on exciton-momentum-dependent relaxation processes, tunable by topological charge. Such interface of vortex beam optics with the centre of mass motion of excitons, influencing dynamics of internal degrees of freedom in 2D materials can be a promising technique to explore novel phenomena in exciton transport and spin-valley physics.

**Methods:**

Monolayers of $WSe_2$ were mechanically exfoliated from commercially available bulk crystals and dry transferred onto a $SiO_2$/Si substrate [37]. An optical vortex beam was generated using an amplitude-only spatial light modulator utilizing computer-generated holograms. Measurements were performed using a closed-cycle optical microscopy cryostat (Montana Instruments) with a variable temperature range of 3.2 K to 295 K. A homebuilt microscope, specialized for sensitive polarization-resolved PL measurement (Supporting Information S1) was utilized for the valley polarization and coherence measurements. A Motorized rotational stage (Thorlabs: Kinesis- KDC101) was used in the polarization-resolved measurements to accurately control the angles of optical elements. Princeton Instruments spectrometer (SP2750) and a liquid nitrogen cooled detector (PyLoN:400BR-eXcelon) were used to measure the PL signal.

**Acknowledgments:**

We acknowledge Sourin Das, Mandar M. Deshmukh, and Chitraleema Chakraborty for their valuable comments on this work. S. D. acknowledges Science and Engineering Research Board (SB/S2/RJN-110/2016, CRG/2018/002845), Ministry of Education STARS (MoE/STARS-1/647), Indian Institute of Technology Kharagpur (IIT/SRIC/ISIRD/2017-2018), Ministry of Human Resource Development (IIT/SRIC/PHY/NTS/2017-18/75) for the funding and support for this work. D. C. acknowledges Council of Scientific and Industrial Research, JRF (09/081(1352)/2019-EMR-I) for the financial assistance.

**Figure: 1**

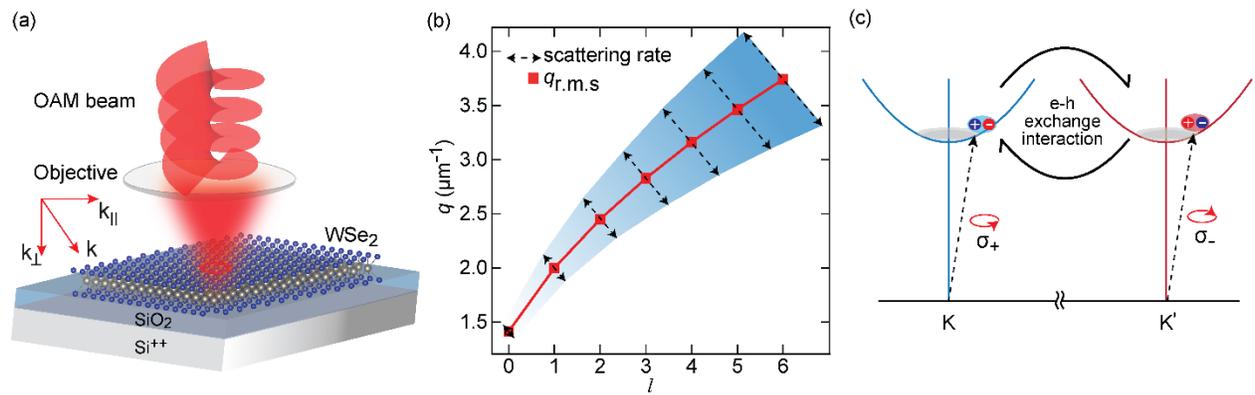

**Figure 1:** (a) Schematic of WSe$_2$ monolayer being excited by OV beam. (b) Increase in exciton momentum ($q_{r.m.s}$) with increase in $l$ number of OV beam; blue gradient region with arrows represents intervalley scattering rate of exciton increasing quadratically with momentum. (c) Diagram showing valley selection rule and the intervalley e-h exchange of bright exciton between $K$ and $K'$ valley.

**Figure: 2**

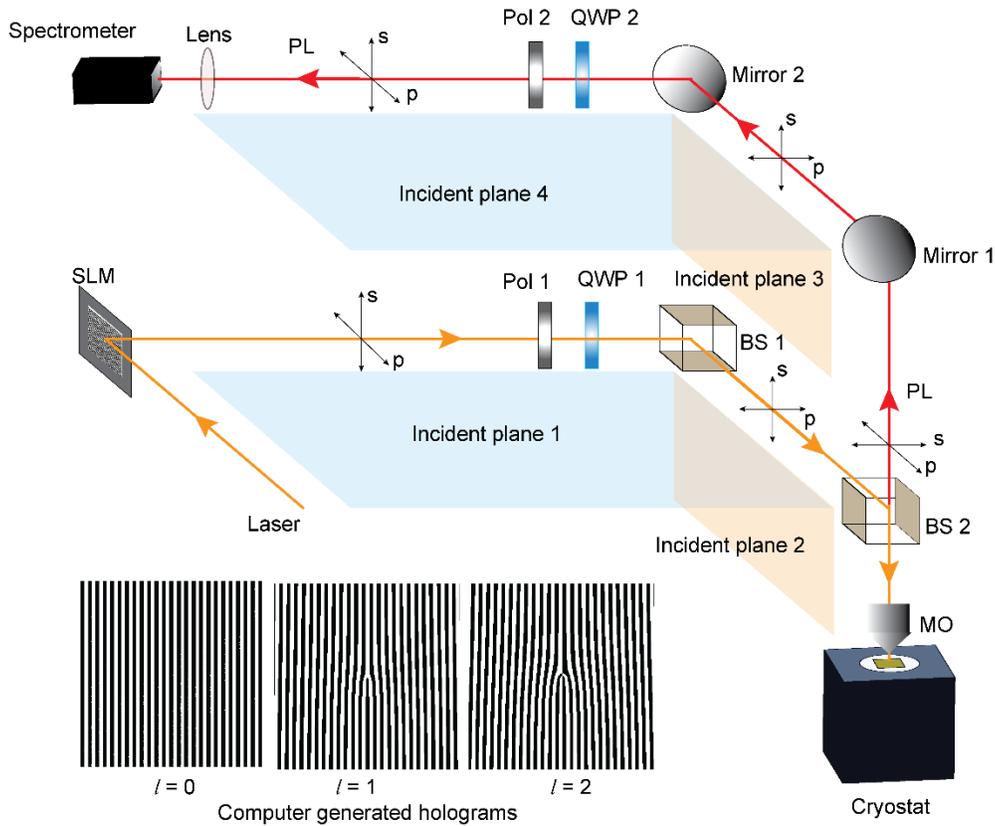

**Figure 2**: Display of amplitude only SLM of computer-generated holograms (CGHs) used to generate optical vortices are shown. To preserve the state of polarization of the excitation beam, two identical beam splitters are used in such a way that the two planes of incidences are orthogonal. A similar arrangement has been made in the collection path of emitted PL by using two identical mirrors. The horizontal and vertical planes of incidences are shown with light-blue and light-orange colours, respectively. A polarizer (Pol 1) and a quarter-wave plate (QWP 1) are placed before the first beam splitter to control the state of polarization of the excitation beam. QWP 2 and Pol 2 are placed after the second mirror (Mirror 2) to analyse the state of polarization of the PL. The sample is placed inside the cryostat, excited through a microscope objective (MO) lens (0.7 NA). The output PL emission is collected through the same MO and is focussed by a lens into a spectrometer slit.

**Figure: 3**

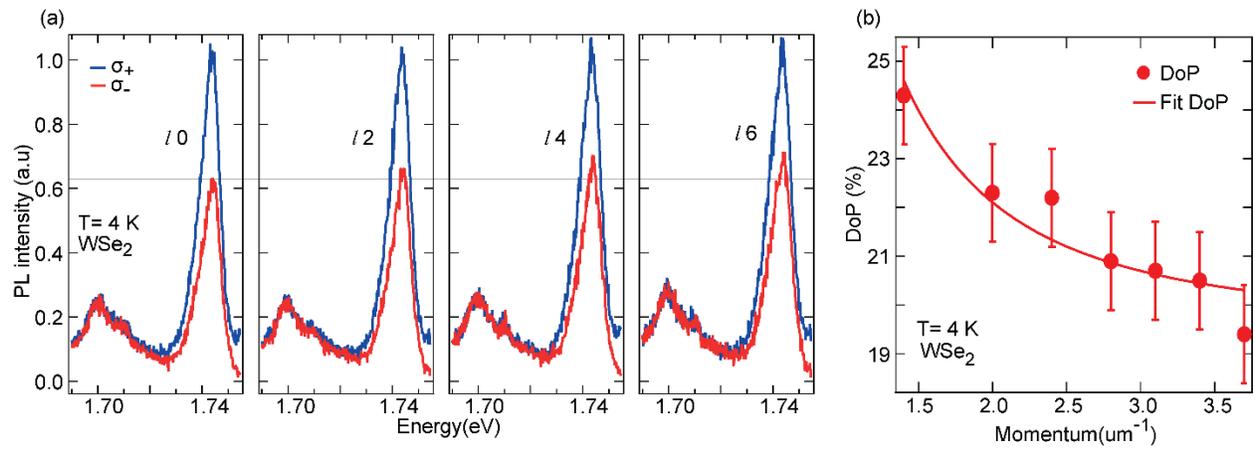

**Figure. 3** (a) Polarization-resolved normalized PL spectra showing valley polarization of excitons at 4 K in $WSe_2$, with an eye guide to visualize the variation of DoP as a function of $l$ varying from 0 to 6. (b) Measured DoP (solid red circles) fitted with the expression obtained from rate equation model (red solid line) are shown.

**Figure: 4**

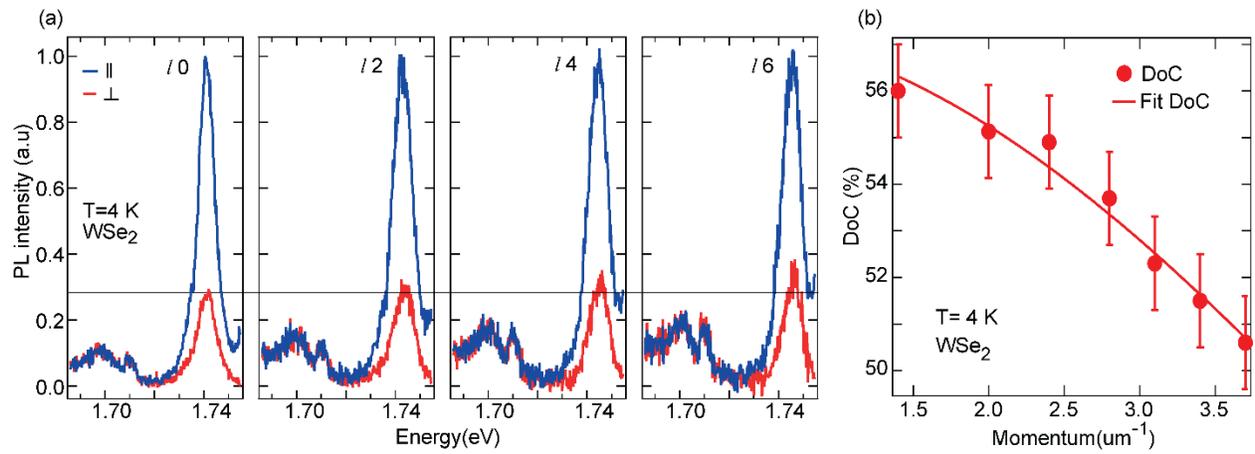

**Figure 4:** (a) Polarization-resolved normalized PL spectra showing valley coherence of excitons at 4 K in WSe$_2$, with an eye guide to visualize the variation of DoC as a function of $l$ varying from 0 to 6. (b) Measured DoC (solid red circles) fitted with the expression obtained from rate equation model (red line) are shown.

**Figure: 5**

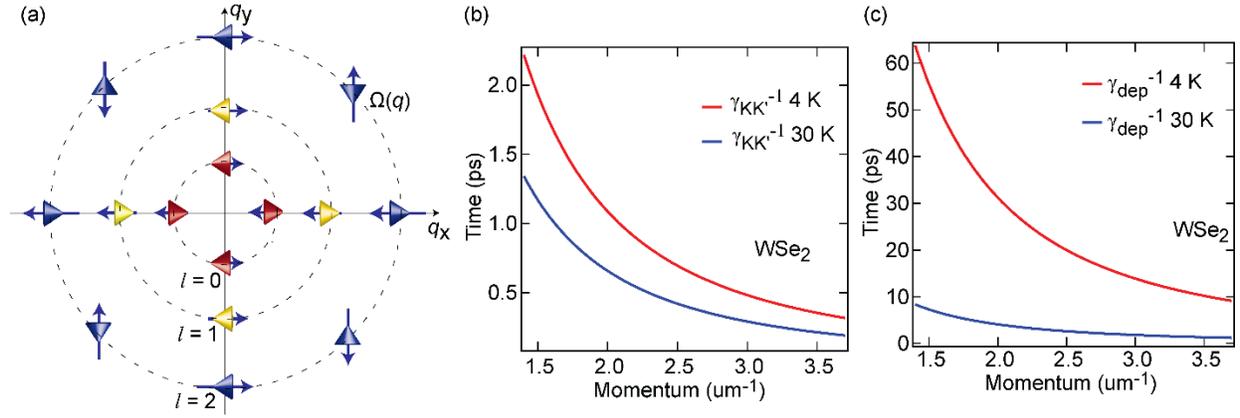

**Figure 5**: (a) Schematic for the visualization of momentum dependent in-plane effective magnetic field $\Omega(q)$ represented in the momentum space (blue arrows) with the precession of excitons pseudospin around $\Omega(q)$ (in colour cones: red, yellow, and blue cones represent precession frequency in increasing order for $l$ varying from 0 to 2). (b) Obtained momentum dependent timescales associated with, relaxation of valley polarization $WSe_2$ monolayer at 4 K and 30 K. (c) Momentum dependent relaxation of valley coherence for $WSe_2$ at 4 K and 30 K.

# Supporting Information

# A steady state approach for studying valley relaxation using optical vortex beam

Aswini Kumar Pattanayak, Pritam Das, Avijit Dhara, Devarshi Chakrabarty, Shreya Paul, Kamal Gurnani, Maruthi Manoj Brundavanam, Sajal Dhara[*]

*Department of Physics, Indian Institute of Technology, Kharagpur-721302, India*

*\* sajaldhara@phy.iitkgp.ac.in*

**S1. Sample preparation and experimental setup :**

Monolayers of $WSe_2$ were mechanically exfoliated from commercially available bulk crystals (HQ Graphene) and dry transferred onto a $SiO_2$/Si substrate[1]. Measurements were performed using a closed-cycle optical microscopy cryostat (Montana Instruments) with a variable temperature range of 3.2 to 295 K. A homebuilt microscope setup, specialized for sensitive polarization-resolved PL measurement was utilized for the valley polarization and coherence measurements as shown in Figure 2 in the manuscript. We have used a 660 nm diode laser (Thor lab: S1FC660) for sample excitation and a Spatial collimator has been used to produce a collimated Gaussian beam. Optical vortex (OV) beams of the desired $l$ are prepared by shining a Gaussian beam on the amplitude only SLM(HOLOEYE LC-R 1080S), which displays the corresponding computer-generated hologram (CGH)[2]. The CGHs for generating $l$ from 0 to 2 have been shown in Figure 2. The 1st order diffracted beam, which contains the OV beam of topological charge $l$, is focused by a microscope objective (MO) of NA 0.7 (60 X) on the monolayer sample. A green LED has been used to image the sample placed inside the cryostat as presented in Figure S1(b).

Valley polarization measurement is performed using a circularly polarized OV beam as the incident excitation, which is prepared by passing the beam through the combination of Pol 1 and QWP 1 as shown in Figure 2. The emitted PL signal is collected through the same MO and is directed into the spectrometer. The circular polarization components of the output PL signal have been measured by using the combination of QWP 2 and Pol 2, as shown in Figure 2 in the manuscript. For valley coherence measurement, linearly polarized light prepared using Pol 1 is used as incident excitation. The plane of polarization of output PL is measured using Pol 2 at pass and cross-orientations. A Motorized rotational stage (Thorlabs: Kinesis- KDC101) was used in the polarization-resolved measurements to accurately control the angles of optical elements. Princeton

Instruments spectrometer (SP2750) and a liquid nitrogen cooled detector (PyLoN:400BR-eXcelon) were used to measure the PL signal.

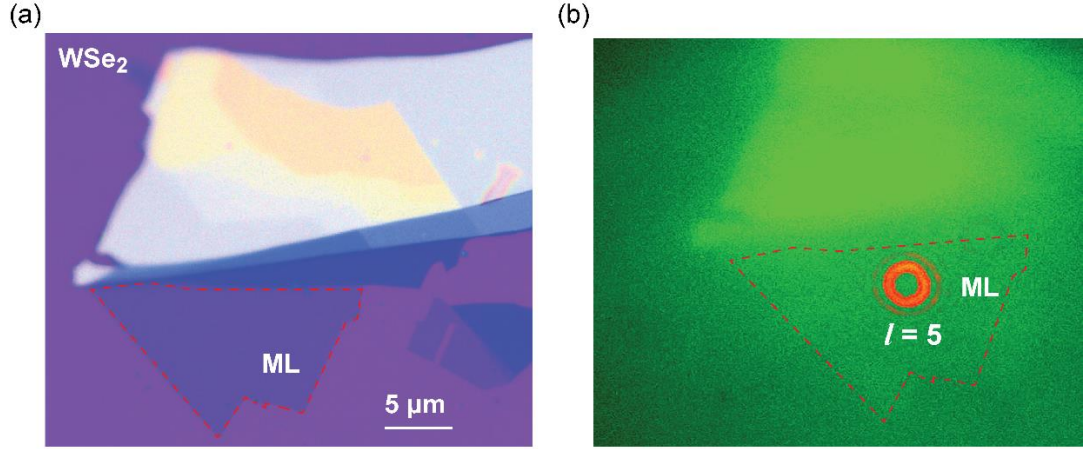

**Figure S1:** (a) SiO$_2$/WSe$_2$ sample under optical microscope. (b) Excitation of WSe2 monolayer by OV beam of $l = 5$.

## S2: In plane momentum distribution of LG beam at the focal plane :

Orbital angular momentum (OAM) and linear momentum are two independent degrees of freedom of light. But, when light is focused via a fixed objective lens (in our case 0.7 N.A), it acquires the in-plane momentum ($k_x \neq 0, k_y \neq 0$) at the focal plane. Vortex beams have different spatial profiles, related to their OAM ($l$). For specific $l$, we have simulated the photon momentum distribution [$I(k_x, k_y)$] at the focal plane of the objective lens as shown in our recent work[3]. We have calculated the root-mean-square momentum ($q_{r.m.s}$) of the distribution and quantify it as an in-plane momentum corresponding to each $l$ number of the vortex beam[3].

## S3: Verification of zero valley coherence and valley polarization at higher temperature with different charges of optical vortex beam:

To verify the polarization sensitivity of our setup we have measured valley polarization and coherence for $WSe_2$ monolayer at 295 K and 200 K for different $l$ as a control case (Fig. S2). The obtained zero DOP and DOC as expected, confirms the absence of any artefacts in our measurement that could have come from the change in $l$ or any rotating parts in the setup.

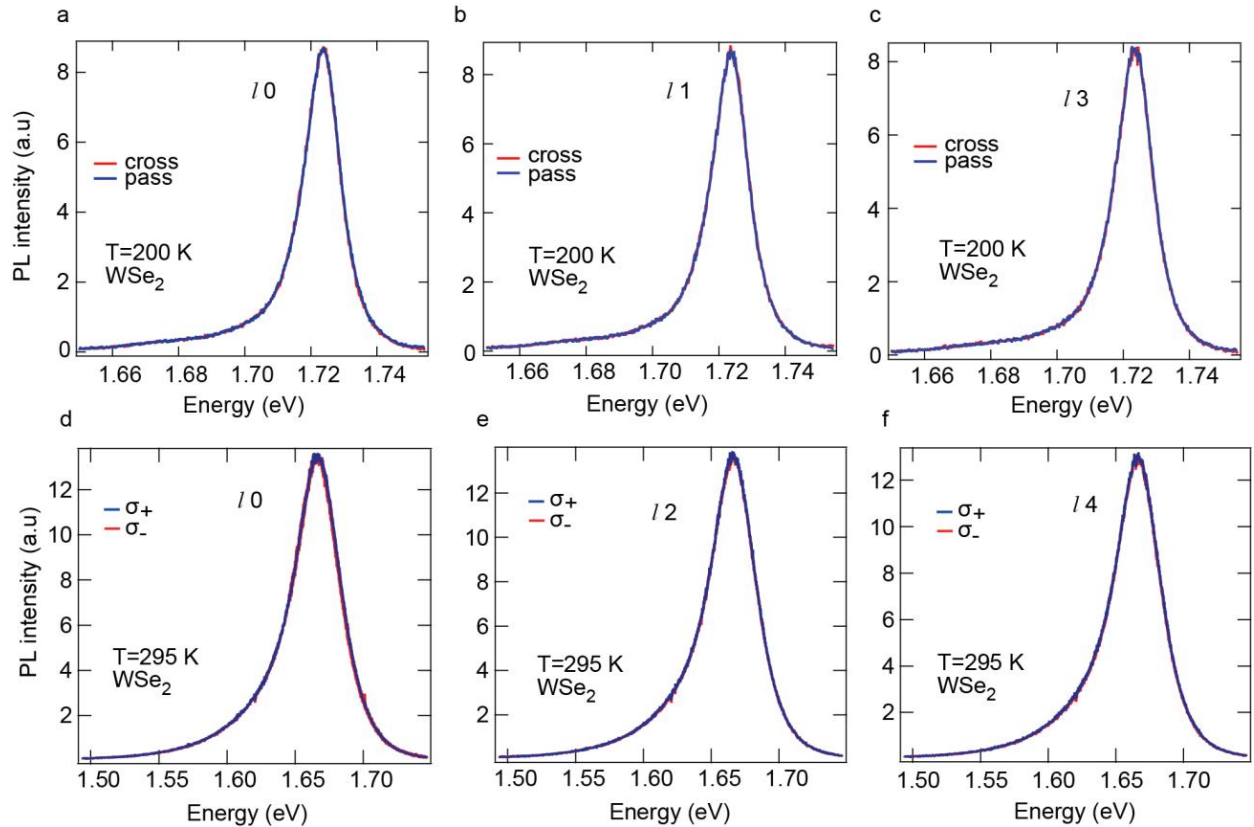

**Figure S3**: **Zero degree of valley polarization and coherence for $WSe_2$ monolayer:** Experimentally obtained polarization-resolved valley coherence (a-c) and valley polarization (d-f) plot for different $l$ at 200 K and 295 K respectively, where zero degree of valley coherence and polarization has been observed.

## S4: Homogeneity of Sample:

Sample quality is an important factor in this experiment as presence of defects not only reduces DoP and DoC but also it reduces the exciton PL emission. For this purpose, we preferred the mechanical exfoliation method from fresh bulk material. Effect of defects in $WSe_2$ is well known and it shows up as a reduction of exciton emission accompanied by a broad PL emission peaks below the exciton emission line (apart from peaks associated with complex exciton species like trions and biexcitons). In Figure S4A, a comparison between sample 1 (used in our experiment) and 2 is shown below to demonstrate the effect of sample quality. It can be observed that sample 1 has comparatively less defect peaks with an intense exciton peak at 10 K. Whereas in sample 2 we see a lot of defect emissions with negligible bright exciton peak at 10 K, serving as an example of a sample with higher defect density. Consequently, the PL spectrum of sample 1 implies a good quality sample, which is conducive to observe the phenomena explained in the manuscript and is similar to the spectrum observed in other studies with high-quality samples[4].

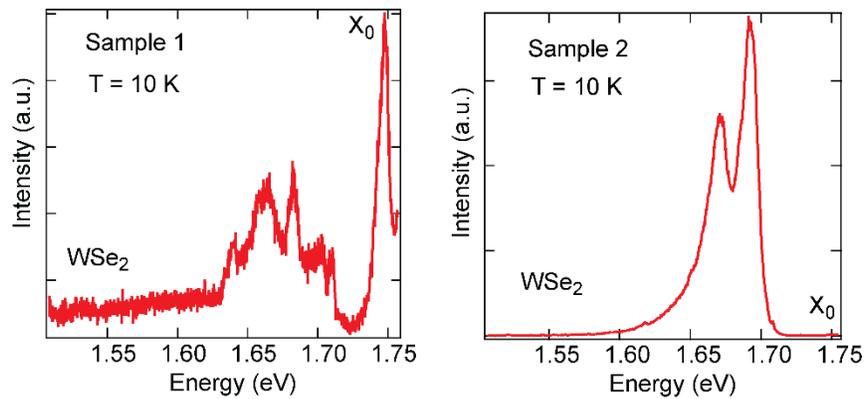

**Figure S4A: Comparison of sample quality**. Sample 1, which has been studied in this manuscript, shows significantly less defect-attributed PL peaks than sample 2, demonstrating its pristine nature which is necessary for minimizing the effect of impurity scattering.

We understand that spatial inhomogeneity of defect density can affect our measurement, therefore, we have carefully studied the position-dependent DoP and have found that sample 1 is mostly homogeneous, as shown in Figure S4B.

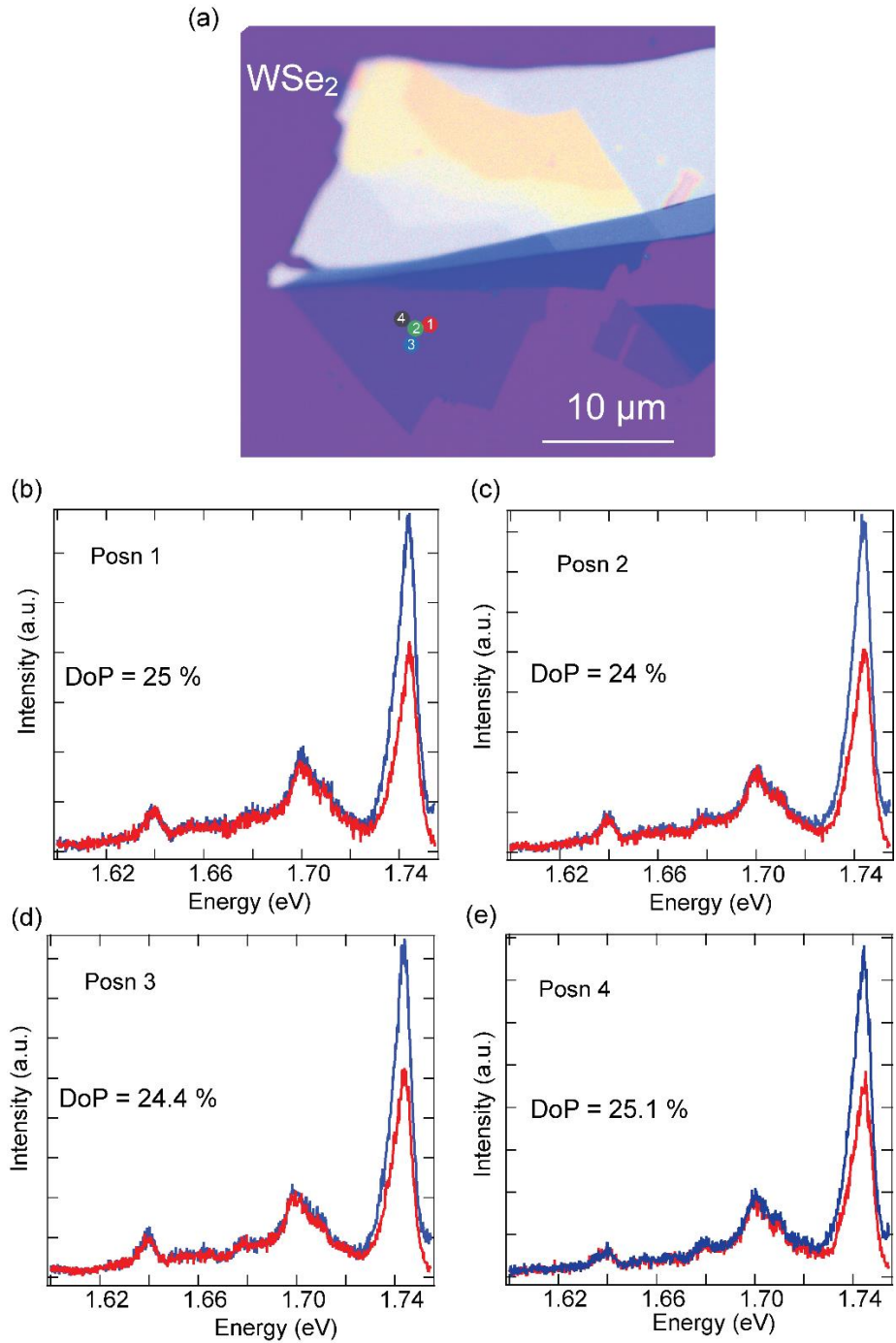

**Figure S4B: (a)** DoP has been measured at positions (1-4) in monolayer WSe$_2$ to verify homogeneity in valley polarization measurement spatially across the sample. **(b-e)** DoP at positions (1-4) has been measured around $(24.5 \pm 1)$%, which remains almost homogeneous spatially.

The energy of bright exciton in WSe$_2$ is around ∼1.747 eV (or 710 nm), whereas the experiment was carried out using a 1.878 eV (660 nm) excitation laser, which can be considered as near-resonant excitation. DoP and DoC decrease with increasing laser detuning. The reduced

DoP and DoC in our case can be attributed to the combined effect of not using exactly resonant excitation as well as the effect of the $SiO_2$ substrate.

## S5. Valley polarization at 30K for OV beams

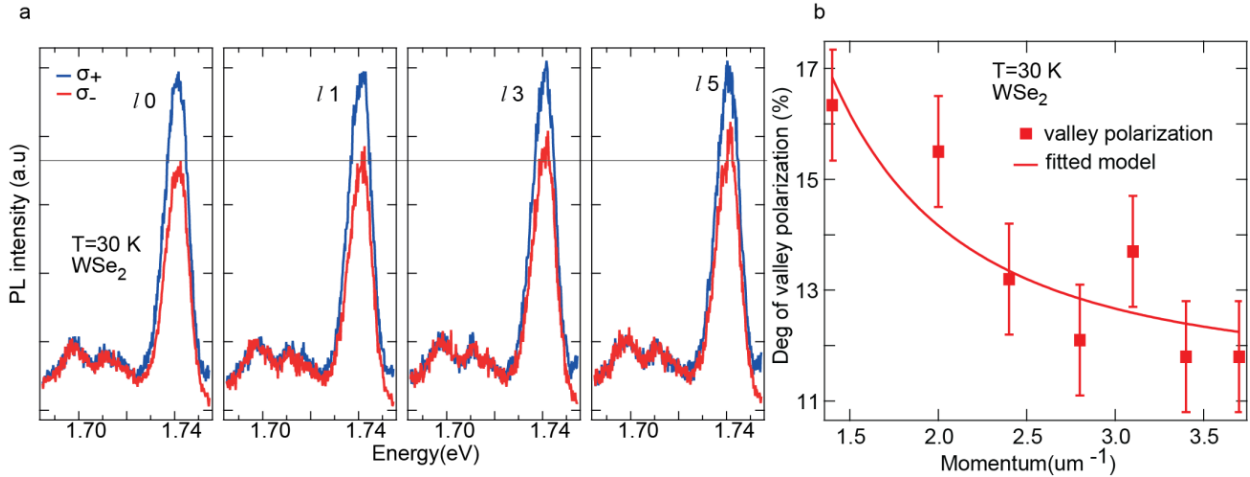

**Figure S5**: **Valley depolarization for different OV beam at 30 K of $WSe_2$ monolayer:** (a) Experimentally obtained helicity resolved PL spectra showing valley polarization for different $l$ at 30 K, where each spectrum is normalized with respect to $\sigma_+$ component of respective $l$. An eye guide has been used to indicate the increase of $\sigma_-$ component, demonstrating the depolarization effect with increasing $l$ of OV beam. (b) Degree of valley polarization has been shown as solid red squares with the fitted red line obtained from the rate equation model.

## S6. Valley coherence at 30K for OV beams

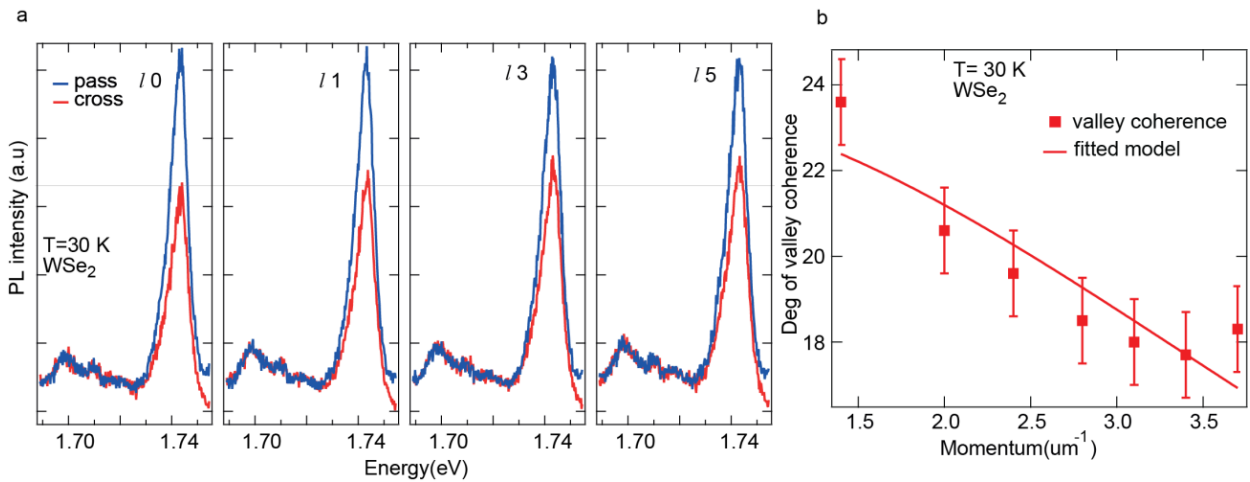

**Figure S6**: **Valley decoherence for different OV beam at 30K of $WSe_2$:** (a) Experimentally obtained polarization-resolved PL spectra showing valley coherence for different $l$ at 30 K, where each spectrum is normalized with respect to pass component of respective $l$. An eye guide has been used to indicate the increase of cross-component, demonstrating the decoherence effect with increasing $l$ of OV beam. (b) Degree of valley coherence has been shown as solid red squares with the fitted red line obtained from the theoretical model.

## S7: Fitted function for valley depolarization and decoherence:

*Fitted function for valley depolarization using rate equation model:*

The solution of bright exciton intensity can be written as,

$$I_{bright} = I_K^b + I_{K'}^b = \frac{2\left(e^{\frac{\Delta E}{K_B T}} g_b \tau_{bd} + (g_b + g_d)\tau_d\right)}{e^{\frac{\Delta E}{K_B T}}(\tau_b + \tau_{bd}) + \tau_d} \quad (1)$$

For valley polarization measurements, as only one valley is excited so we have considered $g_K^b = g_b$, $g_K^d = g_d$, $g_{K'}^b = 0$, $g_{K'}^d = 0$.

Hence from the rate equation model, the degree of polarization can also be obtained as,

$$\rho_c = \frac{1}{1 + \frac{2\tau_b}{\tau_{KK'}} \left[\frac{\tau_{bd} + \tau_d e^{\frac{-\Delta E}{K_B T}}}{(\tau_b + \tau_{bd}) + \tau_d e^{\frac{-\Delta E}{K_B T}}}\right]}$$

For $T = 4\ K$ with $\Delta E = 40\ mev,^5\ e^{\frac{-\Delta E}{K_B T}} \sim 10^{-51}$

$$\rho_c = \frac{1}{1 + \frac{2}{\tau_{KK'}} \frac{\tau_b \tau_{bd}}{\tau_b + \tau_{bd}}}$$

Total population decay of the bright exciton can be defined as, $\frac{1}{\tau_{bright}} = \frac{1}{\tau_b} + \frac{1}{\tau_{bd}}$

$\rho_c = \frac{1}{1 + \frac{2\tau_{bright}}{\tau_{KK'}}}$. Which can also be presented as, $\rho_c = \frac{1}{1 + \frac{2\gamma_{KK'}}{\gamma_{bright}}}$ where $\gamma_{bright} = \gamma_b + \gamma_{bd}$

Here $\gamma_b$ is the radiative recombination rate of the bright exciton and $\gamma_{bd}$ is the bright to dark exciton intravalley scattering rate, $\gamma_{KK'}$ is the intervalley scattering rate from $K$ valley to $K'$ valley.

Since both $\gamma_{bd}$ and $\gamma_{KK'}$ are quadratically depends on $q$, $\rho_c$ can be written as

$$\rho_c = \frac{1}{1 + \frac{2bq^2}{aq^2 + \gamma_b}}$$

where $\gamma_{bd} = aq^2$, $\gamma_{KK'} = bq^2$. Here, a, b is the proportionality constant.[6–8]

$$a = \frac{4\alpha_{so}^2}{\hbar}\left(\frac{\tau^*}{1 + \left(\frac{\Delta_0 \tau^*}{\hbar}\right)^2}\right), \quad b = \left(\sqrt{5}C\alpha(1)\hbar\right)^2 \tau^*$$

Here $\alpha_{so}$, $\tau^*$, $\Delta_0$ are Rashba spin-orbit coupling parameter, exciton elastic momentum scattering time, and exchange splitting energy between bright and dark states. C, $\alpha(1)$ are the material parameters.

Considering $F = \frac{2bq^2}{aq^2+\gamma_b}$, $\frac{dF}{dq} = \frac{4b\gamma_b q}{(aq^2+\gamma_b)^2} > 0$ for $(a, b, \gamma_b) > 0$

It indicates that $F$ is an increasing function with increasing $q$, leading to valley depolarization. The physical explanation for the depolarization process has been described in the manuscript.

*Fitted function for valley decoherence:*

The degree of valley coherence can be written as[9,10],

$$\rho_l = \frac{1}{\left(1 + \frac{\gamma_v + \gamma_{dep}}{\gamma_x}\right)}$$

Where, $\gamma_v$ is the incoherent intervalley scattering rate, which is proportional to the phonon population in the system and $\gamma_{dep}$ is the pure dephasing rate and $\gamma_x$ is the exciton relaxation rate.

Since $\gamma_{dep}$ is quadratically depends on $q$, $\rho_l$ can be written as

$$\rho_l = \frac{1}{1 + \frac{\gamma_v + cq^2}{\gamma_x}} \quad (4)$$

where $\gamma_{dep} = cq^2$. Here, $c = k\left(\sqrt{5}C\alpha(1)\hbar\right)^2 \tau^*$ is the proportionality constant.

From the fitting of valley polarization and coherence data for OV beams the obtained momentum dependent valley dynamics timescale for 4K and 30K has been presented in manuscript Fig. 5(b-c)

## S8: Exciton momentum dependence of intravalley scattering rate

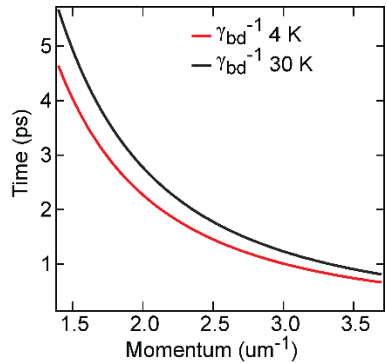

**Figure S8:** Exciton momentum dependent timescales associated with intravalley scattering rate for WSe$_2$ monolayer at 4 K and 30 K.